\documentclass[twocolumn,showpacs,preprintnumbers,amsmath,amssymb,aps,prb,superscriptaddress]{revtex4-1}

\pdfoutput=1
\usepackage[pdftex]{graphicx}
\usepackage[english]{babel}

\usepackage{cleveref}

\usepackage{dcolumn}% Align table columns on decimal point\right] 
\usepackage{bm}% bold math

\usepackage{color}
\usepackage{amstext}
\usepackage{braket}

\begin{document}

%%%%%%%%%%%%%%%%%%%%%%%%%%%%%%%%%%%%%%%%%%%%%%%%%%%%%%%%%%%%%%%%%%%%%%%%%%
%                               Title                                    %
%%%%%%%%%%%%%%%%%%%%%%%%%%%%%%%%%%%%%%%%%%%%%%%%%%%%%%%%%%%%%%%%%%%%%%%%%%

\title{Edge currents in frustrated Josephson junction ladders}

\author{A. M. Marques}
\email{anselmomagalhaes@ua.pt}
\affiliation{Department of Physics $\&$ I3N, University of Aveiro, 3810-193 Aveiro, Portugal}
\author{ F. D. R. Santos}
\affiliation{Department of Physics $\&$ I3N, University of Aveiro, 3810-193 Aveiro, Portugal}
\author{ R. G. Dias}
\affiliation{Department of Physics $\&$ I3N, University of Aveiro, 3810-193 Aveiro, Portugal}

\date{\today}

%%%%%%%%%%%%%%%%%%%%%%%%%%%%%%%%%%%%%%%%%%%%%%%%%%%%%%%%%%%%%%%%%%%%%%%%%%
%                              abstract                                  %
%%%%%%%%%%%%%%%%%%%%%%%%%%%%%%%%%%%%%%%%%%%%%%%%%%%%%%%%%%%%%%%%%%%%%%%%%%

\begin{abstract}
We present a numerical study of quasi-1D frustrated Josephson junction ladders with diagonal couplings and open boundary conditions, in the large capacitance limit. 
We derive a correspondence between the energy of this Josephson junction ladder and the expectation value of the Hamiltonian of an analogous tight-binding model, and show how the overall superconducting state of the chain is equivalent to the minimum energy state of the tight-binding model in the subspace of one-particle states with uniform density. 
To satisfy the constraint of uniform density, the superconducting state of the ladder is written as a linear combination of the allowed $k$-states of the tight-binding model with open boundaries. 
Above a critical value of the parameter $t$ (ratio between the intra-rung and inter-rung Josephson couplings), the ladder spontaneously develop currents at the edges which spread to the bulk as $t$ is increased until complete coverage is reached. 
Above a certain value of $t$, which varies with ladder size ($t=1$ for an infinite-sized ladder), the edge currents are destroyed. The value $t=1$ corresponds, in the tight-binding model, to the opening of a gap between two bands. 
We argue that the disappearance of the edge currents with this gap opening is not coincidental, and that this points to a topological origin for these edge current states.
\end{abstract}

\pacs{74.25.Dw,74.25.Bt}

\maketitle
%%%%%%%%%%%%%%%%%%%%%%%%%%%%%%%%%%%%%%%%%%%%%%%%%%%%%%%%%%%%%%%%%%%%%%%%%%
%                              body of paper                             %
%%%%%%%%%%%%%%%%%%%%%%%%%%%%%%%%%%%%%%%%%%%%%%%%%%%%%%%%%%%%%%%%%%%%%%%%%%
%\section{Introduction}

\section{Introduction}
\label{sec:intro}
In solid-state systems, the bulk-edge correspondence states that, if the bulk has non-trivial values for quantities that are associated with its topology, such as the Chern number or the $\mathbb{Z}_{2}$ invariant, then there must be topologically protected states localized at the edges \cite{Hasan2010,Shen2012}. 
Different topological states have been proposed to be present at the edges of Q1D (quasi-1D), 2D and 3D materials \cite{Hasan2010,Shen2012}.
The Su-Schrieffer-Heeger (SSH) model \cite{Su1979} is a tight-binding (TB) model that provides a simple example of a Q1D system where, due to a non-trivial bulk winding number in certain regions of the parameter space, one has zero-energy states with an electronic probability density mainly distributed at the edge and with a decaying tail to the bulk.

Some theoretical models predict the observation of analogous topological effects in Q1D systems with superconducting pairing terms. One example is the existence of Majorana fermions localized at the ends of semiconductor nanowires coupled to superconductors (good reviews on the subject are available \cite{Alicea2012,Leijnse2012,Beenakker2013}). 
In the context of Josephson junction (JJ) chains, there has been a recent proposal for the observation of a new kind of Majorana fermions at the ends of a system of three coupled JJ ladders \cite{Pino2015}.

In this paper, we present a numerical study of open boundary JJ ladders. 
In the large capacitance limit (classical limit), the minimum energy solution of a periodic JJ chain can be found from, and written in terms of, a single one-particle $k$-state with uniform density (a Bloch state) of an analogous TB system, by obeying some simple correspondence rules \cite{Choi1985,Yi1997,Dias2014}. 
If one considers an open JJ chain consisting of an array of equal superconducting condensates, each with its own superconducting phase, the overall superconducting state will be written as a linear combination of the allowed open TB chain $k$-states (which individually do not have uniform density) in order to respect the constraint of uniform density\ \cite{Dias2014}. 
%The one-particle states in TB systems with open boundaries do not have uniform density. 
%But when writing the Hamiltonian of the JJ array in the TB formulation, an exact correspondence can only be preserved if one imposes the constraint that the superconducting state should have uniform density. 
So in order to find the superconducting state of the JJ chain, one searches for the minimum energy state within the subspace of one-particle states with uniform density of the corresponding TB chain with open boundaries.
This implies that the superconducting state cannot be localized, at least in the usual sense, and by this we mean as the analog of a localized electronic wave function with a decaying probability density.

However, even if we are constrained to uniform density states, we can pose a new question: can we have states that show localized behavior in the derivative of the quantum wave function?
More precisely, in the context of JJ ladders with open boundary conditions, is it possible to have a JJ ladder state characterized by a decaying edge current density, given that the probability density is uniform?
And, if so, how should it manifest? 
We argue that, if present, these new "localized current states" should translate into experimentally detectable changes in the behavior of the superconducting phases configuration. 
The quantum current operator, which implies a derivative of the wave function, can be finite even if the wave function has uniform density, due to the existence of a phase gradient.

In our JJ-TB correspondence, the phase of each site is interpreted as the superconducting phase of the respective superconducting island, and Josephson currents appear as a consequence of finite superconducting phase differences between coupled superconducting islands.
In order for these finite superconducting phase differences to develop spontaneously the JJ array has to be frustrated.
The topic of frustrated JJ arrays has been addressed since the late seventies \cite{Villain:1977wb,Shih1984,Halsey1985}. Recently, there has been a renewed interest in JJ arrays \cite{Douccot2002,Protopopov2004,Protopopov2006,Pop2008,Pop2010} due to, among others, two particular reasons: the direct analogy one can establish between these and frustrated models of classical XY spin chains \cite{Lee1984,Choi1985,KAWAMURA:1988vk}, when the temperature of the JJ chain is sufficiently low; and the theoretical proposal of using JJ chains as the basic elements in quantum computation \cite{Pop2008,Gladchenko2009,Pop2010}.

Recently it has been shown that frustrated multiband superconductors display chiral order parameters (with a superconducting phase configuration that breaks time-reversal symmetry \cite{Tanaka2001,Tanaka2002,NgT.K.2009,Stanev2010,Tanaka2010,Dias2011,Garaud2011,Hu2012,Wilson2013,Orlova2013,Gillis2014}).
Some possible ways of experimentally observing and identifying these chiral solutions have already been proposed \cite{Garaud2014,Stanev2015,Marques2015,Yerin2015}.
The interband interaction between different condensates in a multiband superconductor has been shown, both theoretically \cite{Leggett1966,Agterberg2002} and experimentally \cite{Moll2014}, to be analogous to an intrinsic JJ between distinct superconductors. 
In this analogy, a repulsive interband interaction, of the kind thought to be present in sign-reversed two-band superconductors \cite{Kuroki2008,Mazin2008a}, is formally the same as a $\pi$-junction between two superconducting islands. Recent experimental evidence, based on the observation of in-gap states in iron-based superconductors with non-magnetic impurities, supports the existence of a sign reversal between the superconducting phases of different bands of a multiband superconductor \cite{Zhang2014,Mizukami2014}.
A JJ array becomes frustrated if an odd number of repulsive interactions is present in a unit cell (note that frustration can also be introduced in a JJ array by an external magnetic field). 
This can be achieved in two ways: by explicitly introducing $\pi$-junctions or, following the aforementioned analogy, introducing sign-reversed two-band superconductors instead of the $\pi$-junctions \cite{Dias2014}. Either way the results are the same for both cases.

We perform in this paper numerical studies to characterize frustrated JJ ladders with diagonal couplings and open boundaries. 
We have found that, above a critical value of the ratio between the intra-rung and inter-rung Josephson couplings, edge currents start to develop at the outermost squares of the ladders and propagate to the bulk as this ratio is increased, and disappear after these edge currents extend to the whole ladder. We explore the JJ $\leftrightarrow$ TB correspondence and show how one can decompose the JJ ladder state in the allowed $k$-states of the bands of the TB model, and relate these decompositions, for different values of the ratio, to their respective current configurations.
From the results obtained, which show that the disappearance of edge currents occurs in parallel with a gap opening between the TB bands, we suggest that these two features are likely to be related, and not independent, which could reflect a new topological origin for these edge current states.

The remaining part of this paper is organized in the following way. 
In section \ref{sec:jj_tb}, we define our model Hamiltonian for a frustrated JJ ladder and show how to translate it into the corresponding TB model and how to write the superconducting state of the JJ array as a state of this TB model. 
Section \ref{sec:open_boundaries} addresses the case of open JJ ladders. 
The phase diagram  of the chiral edge currents versus the relative strength of the Josephson couplings is obtained. 
For different values of this relative strength, we analyze the decomposition of the superconducting state in terms of $k$-states the TB chain. 
We end section \ref{sec:open_boundaries} addressing how the edge currents are dependent on the condition of perfect flat bands.
Finally, we conclude in section \ref{sec:conclusions}.

%%%%%%%%%%%%%%%%%%%%%%%%%%%%%%%%%%%%%%%%%%%%%%%%%%%%%%%%%%%%%%%%%%%%%%%%%%%%%%%%%%% SECTION %%%%%%%%%%%%%%%%%%%%%%%%%%%%%%%%%%%%%%%%%%%%%%%%%%%%%%%%%%%%%%%%%%%%%%%%%%%%%%%%%%%%%%%%%%%%%%%%%%%%
\section{JJ ladder and the TB correspondence}
\label{sec:jj_tb}

\begin{figure*}[t]
\begin{center}
\includegraphics[width=0.75 \textwidth]{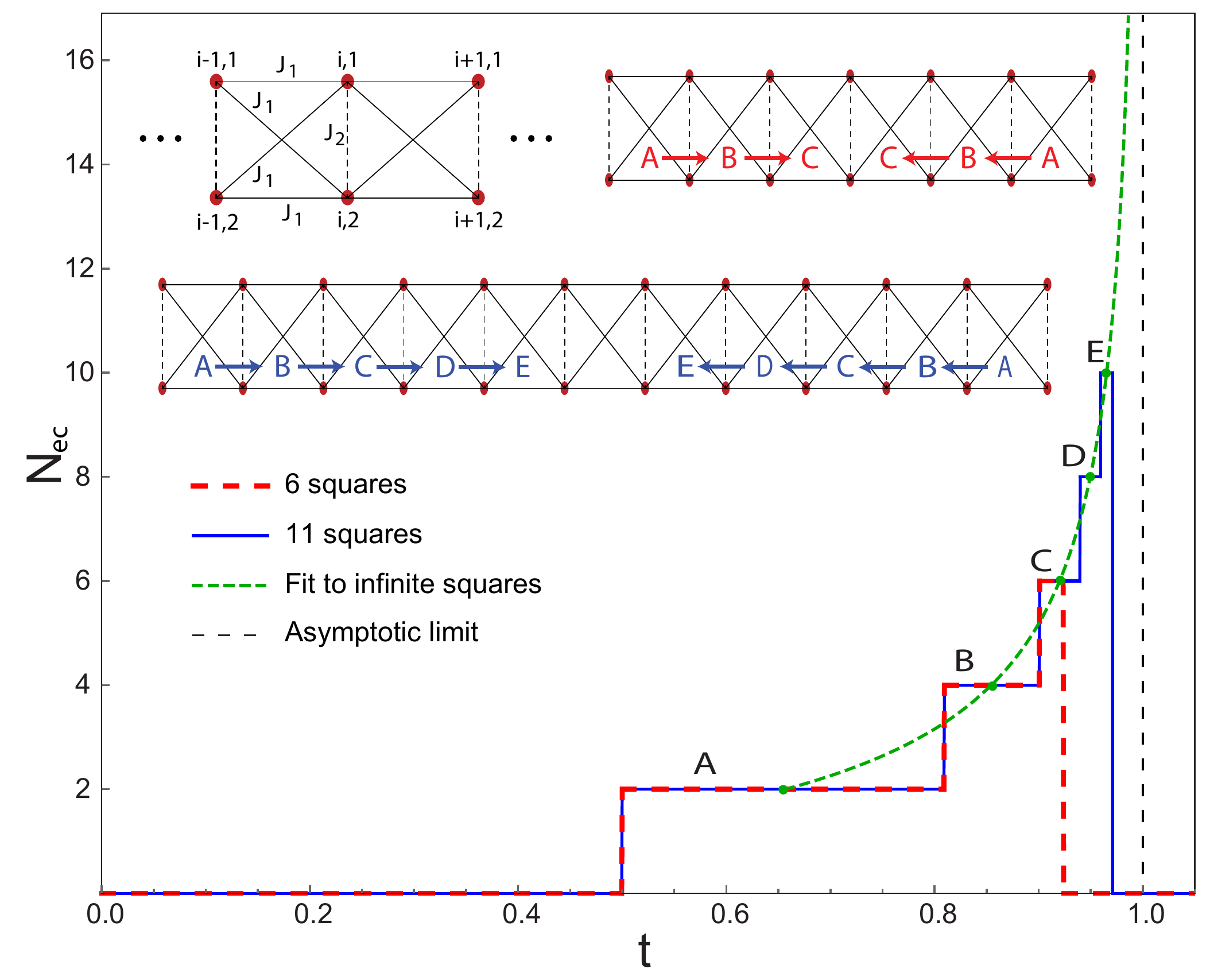}
\end{center}
\caption{Number $N_{ec}$ of squares in the chain with edge currents as a function of $t$, the relative strength of the hopping constants defined in (\ref{eq:hamilt_kernel}). In region A edge currents develop at the outermost squares, which successively propagate to adjacent innermost squares as $t$ is increased, while \textit{persisting} in the previous squares- regions B-C and B-E for the chains with 6 and 11 squares, respectively. So, for instance, in region B (where $N_{ec}=4$) the edge currents propagate to the squares labeled B, but persist in those labeled A, and so on. A further increase in $t$ destroys the edge current states (the $t$ intervals above the sudden drops in the red dashed and blue solid curves). Green dashed curve represents an extrapolation to an infinite-sized open chain, according to the fitting expression $a/\sqrt{1-t}+b$, with $a=2.184$ and $b=-1.729$, which has an asymptote at $t=1$ (vertical black dashed line). The fitting agrees very well with our numerical result for 11 squares.}
\label{fig:N_t_phase_diagram}
\end{figure*}

We start by considering an infinite and periodical Q1D JJ ladder with diagonal couplings in the large capacitance limit (see the diagram at the upper left corner of Fig.~\ref{fig:N_t_phase_diagram}) with energy given by
%\begin{equation}
%\label{eq:hamiltonian_JJ}
%H=\sum_j^N \sum_{l,m=1}^2 \Big[J_1\cos(\phi_{j,l}-\phi_{j+1,m}) + J_2 \cos(\phi_{j,1}-\phi_{j,2}) \Big],
%\end{equation}
\begin{eqnarray}
H&=&H_1+H_2,
\label{eq:hamiltonian_JJ}
\\
H_1&=&\sum_j^N \sum_{l,m=1}^2J_1\cos(\phi_{j,l}-\phi_{j+1,m}),
\label{eq:hamiltonian_JJ1}
\\
H_2&=&\sum_j^N J_2 \cos(\phi_{j,1}-\phi_{j,2}),
\label{eq:hamiltonian_JJ2}
\end{eqnarray}
where $N+1 \to 1$, $J_1<0$ is the inter-rung Josephson coupling, $J_2>0$ is the intra-rung Josephson coupling that introduces frustration and which can be either a $\pi$-JJ between the two superconducting islands of a unit cell or a repulsive interband coupling between the bands of a two-band superconductor (one has a linear JJ chain in this case). 
We point out that similar frustrated geometries have been considered in the context of spin ladders, where different spin exchange interactions take the place of the Josephson couplings \cite{Honecker2000,Fouet2006,Verkholyak2012,Chen2016}.
A local Josephson or interband current is defined, in energy units, as the derivative of (\ref{eq:hamiltonian_JJ1}) and (\ref{eq:hamiltonian_JJ2}) with respect to the phases,
\begin{eqnarray}
j_{i,l\to i+1,m}&=&\frac{\partial H_1}{\partial \phi_{i+1,m}}=J_1 \sin(\phi_{i,l}-\phi_{i+1,m}),
\label{eq:current_inter}
\\ 
j_{i,1\to i,2}&=&\frac{\partial H_2}{\partial \phi_{i,2}}=J_2 \sin(\phi_{i,1}-\phi_{i,2}),
\label{eq:current_intra}
\end{eqnarray}
with $j_{a\to b}=-j_{b\to a}$. Given that $j\propto \sin(\Delta\phi)$, where $\Delta\phi$ is a superconducting phase difference, currents are present when $\Delta\phi\neq 0,\pm\pi$, which is only possible if the system is frustrated.

%%%%%%%%%%%%%%%%%%%%%%%%%%%%%%%%%%%%%%%%%%%%%%%%%%%%%%%%%%%%%%%%%%%%%%%%%%%%%%%%%%% FIGURE %%%%%%%%%%%%%%%%%%%%%%%%%%%%%%%%%%%%%%%%%%%%%%%%%%%%%%%%%%%%%%%%%%%%%%%%%%%%%%%%%%%%%%%%%%%%%%%%%%%%%
The energy expression (\ref{eq:hamiltonian_JJ}) can be interpreted as the expectation value of the energy of a one-particle state with uniform density in a TB system. In second quantization formalism the JJ array state can be written as the following one-particle state
\begin{equation}
\label{eq:sc_state}
\vert \psi \rangle = \sum_{j=1}^N \sum_{l=1}^2 e^{i \phi_{j,l}} c^\dagger_{j,l} \vert \emptyset \rangle,
\end{equation}
where $\vert \emptyset \rangle$ is the vacuum state, $c^\dagger_{j,l}$ is the creation operator at site ($j,l$) and $\phi_{j,l}$, in our JJ $\leftrightarrow$ TB analogy, is the superconducting phase of the superconducting island ($j,l$) (or of one of the bands of a two-band superconductor). 
This $\vert \psi \rangle$ state is not normalized, in order to preserve our JJ chain $\leftrightarrow$ TB chain correspondence. Using this notation the TB Hamiltonian corresponding to expression (\ref{eq:hamiltonian_JJ}) becomes
\begin{equation}
\label{eq:hamilt_tb}
H_{TB}=\frac{1}{2}\sum_j^N \sum_{l,m=1}^2 \Big[-t_1 c^\dagger_{j,l}c_{j+1,m} +t_2 c^\dagger_{j,1}c_{j,2} + H.c. \Big] ,
\end{equation}
where the following substitutions, $J_1 \to -t_1$ and $J_2 \to -e^{i \pi}t_2=t_2$, with $t_1,t_2>0$, are implied. 
Now $J_i$ is reinterpreted as a hopping parameter with $J_2$ carrying a $\pi$ Peierls phase between the two sites of the unit cell, as illustrated in Fig.~\ref{fig:bandscurrentsplot}(a). 
By computing $\langle \psi \vert H_{TB} \vert \psi \rangle$ with the above correspondences  one recovers the energy expression (\ref{eq:hamiltonian_JJ}). 
We change the representation to the $k$-space with the Fourier transforms $c^\dagger_{j,l}=\frac{1}{\sqrt{N}}\sum_k e^{-i kj}c^\dagger_{k,l}$ and $c_{j,l}=\frac{1}{\sqrt{N}}\sum_k e^{i kj}c_{k,l}$, where $N$ is the number of unit cells and $k=\frac{2\pi}{N}n$ with $n=0,1,\dots,N-1$. 
In this representation the TB Hamiltonian becomes
\begin{align}
    H_{TB} &= \frac{1}{2}\sum_k (c^\dagger_{k,1}\  c^\dagger_{k,2})\mathcal{H}(k)\left(\begin{matrix}
           c_{k,1} \\
           c_{k,2}
         \end{matrix}\right),
\end{align}
where the kernel is given by
\begin{equation}
\label{eq:hamilt_kernel}
\frac{\mathcal{H}(k)}{2t_1} = \left(\begin{matrix}
           -\cos(k) & -\cos(k)+t \\
           -\cos(k)+t & -\cos(k)
         \end{matrix}\right) ,
\end{equation}
where $t=\frac{t_2}{2t_1}$, the relative strength of the hopping/JJ-coupling constants, becomes the only relevant parameter.
%, and so $2t_1$ is set to 1 for simplicity hereafter. 
$\mathcal{H}(k)$ can be readily diagonalized and $H_{TB}$ becomes
\begin{equation}
\label{eq:hamilt_tb_diagon}
H_{TB}=t_1\sum_k \Big[\varepsilon_+ a^\dagger_k a_k + \varepsilon_- b^\dagger_k b_k \Big] ,
\end{equation}
with $\varepsilon_+=t-2\cos(k)$ (itinerant band), $\varepsilon_-=-t$ (flat band), $a^\dagger_k=\frac{1}{\sqrt{2}}(c^\dagger_{k,1} + c^\dagger_{k,2})$ and $b^\dagger_k=\frac{1}{\sqrt{2}}(c^\dagger_{k,1} - c^\dagger_{k,2}$).

%%%%%%%%%%%%%%%%%%%%%%%%%%%%%%%%%%%%%%%%%%%%%%%%%%%%%%%%%%%%%%%%%%%%%%%%%%%%%%%%%%% SECTION %%%%%%%%%%%%%%%%%%%%%%%%%%%%%%%%%%%%%%%%%%%%%%%%%%%%%%%%%%%%%%%%%%%%%%%%%%%%%%%%%%%%%%%%%%%%%%%%%%%%
\section{Ladders with open boundaries}
\label{sec:open_boundaries}

The knowledge of the TB eigenstates of the periodic ring described in the previous section allows one to diagonalize the same Hamiltonian with open boundary conditions.
 We follow the usual recipe of treating an open chain of $N$ unit cells as a periodic chain of $N+2$ unit cells with the condition of zero wavefunction amplitude at the extra sites, accomplished by combining symmetric $k$-states,
\begin{equation}
c^\dagger_{k,1,open}= \frac{1}{\sqrt{2}}(c^\dagger_{k,1}-c^\dagger_{-k,1})= \sqrt{\frac{2}{N+1}}\sum_{j=1}^N\sin(kj)c^\dagger_{j,1},
\end{equation}
where $k=\frac{\pi}{N+1}n$ with $n=1,2,\dots,N$. The same procedure is followed for $c^\dagger_{k,2,open}$. 
We write $a^\dagger_k$ and $b^\dagger_{k}$ as the same linear combination of the new operators of the open chain as in (\ref{eq:hamilt_tb_diagon}). 
The JJ array state $\vert \psi \rangle$ in (\ref{eq:sc_state}) becomes, in terms of $a^\dagger_k$ and $b^\dagger_k$,
\begin{eqnarray}
\vert \psi \rangle &=& \sqrt{\frac{1}{N+1}}\sum_k \big[\alpha_{k_+} a^\dagger_k + \alpha_{k_-} b^\dagger_k \big]\vert \emptyset \rangle,
\label{eq:psi_tb}
\\
\alpha_{k_\pm} &=& \sum_j \sin(kj)\big[ e^{i\phi_{j,1}} \pm e^{i\phi_{j,2}} \big],
\label{eq:coefficients}
\end{eqnarray}
so the $\vert \psi \rangle$ state is decomposed in $k$-states belonging to the two bands of the TB model, with coefficients $\alpha_{k_\pm}$. 
The total energy of the chain is just the expectation value of the Hamiltonian,
\begin{equation}
\label{eq:energy_tb}
E=\langle \psi \vert H_{TB}\vert \psi \rangle = \frac{t_1}{N+1} \sum_k \Big(\vert \alpha_{k_+}\vert^2 \varepsilon_+ + \vert \alpha_{k_-}\vert^2 \varepsilon_-\Big),
\end{equation}
which has to be equal to the value computed from (\ref{eq:hamiltonian_JJ}). 

We show below that a variation in the parameter $t$ leads to the appearance of edge currents in the JJ array and the evolution of these currents implies different decompositions of the JJ ladder state in terms of the allowed $k$-states of the corresponding TB model. The procedure we adopt is as follows: i) we first minimize numerically the energy (\ref{eq:hamiltonian_JJ}) with respect to the superconducting phases in order to find the minimum energy configuration; ii) after finding the full set of superconducting phases, we then calculate the coefficients $\alpha_{k_\pm}$ in (\ref{eq:coefficients}); iii) we calculate the energy of the TB model by (\ref{eq:energy_tb}) and check that it is consistent with the minimum energy solution found in (\ref{eq:hamiltonian_JJ}); iv) finally the value of each current is calculated using (\ref{eq:current_inter}) and (\ref{eq:current_intra}).

The $N_{ec}$ vs. $t$ phase diagram, where $N_{ec}$ is the number of squares in a ladder with edge currents, is shown in Fig.~\ref{fig:N_t_phase_diagram} for the cases of ladders with 6 and 11 squares. 
In the interval $t\in[0,0.5[$ neither of the ladders has edge currents. 
At $t=0.5$, both ladders enter in region A of the phase diagram where edge currents develop at the outermost squares, also labeled A at their representation at the top of Fig.~\ref{fig:N_t_phase_diagram}. 
Above $t=0.809$ (in region B), the edge currents penetrate further into the ladder, covering now the first and second outermost squares, labeled A and B respectively. 
Above $t=0.905$ one enters region C where the edge currents further penetrate into the ladder, occupying squares A-B-C which, for the case of 6 squares, already corresponds to the whole ladder. 

In the case of the ladder with 6 squares, the edge currents, with increasing $t$, are eventually destroyed at the right edge of region C, as reflected by the sudden drop in the red dashed curve in Fig.~\ref{fig:N_t_phase_diagram}.
The case where one has a perfectly flat band allows the existence of decoupled currents within each square above this sudden drop [see Fig.~\ref{fig:bandscurrentsplot}(f)]. However, the broadening of the flat band drives these currents to zero [see Fig.~\ref{fig:bandscurrentsplotiten}(f)]. 
The chain with 11 squares is long enough to allow further propagation of the edge currents, and one has additional regions D and E, where the edge currents also circulate within the respective D and E squares of the ladder. One should keep in mind that when moving from a region to the next, the edge currents \textit{persist} in the corresponding squares of the previous region. 
When the number of squares in a chain is odd, the edge currents do not reach the central square (since all states, in particular those with edge currents, should be symmetric with respect to the center of the ladder), so full coverage, in the chain with 11 squares, occurs in region E of the phase diagram of Fig.~\ref{fig:N_t_phase_diagram}. 
The disappearance of the edge currents occurs at the right edge of region E, where the sudden drop in the blue solid curve is observed in Fig.~\ref{fig:N_t_phase_diagram}.

In Fig.~\ref{fig:N_t_phase_diagram}, we have introduced a fitting curve giving the centers of the plateaus for an open ladder in the limit of infinite squares (green dashed curve). 
The fitting expression is $a/\sqrt{1-t}+b$ with $a=2.184$ and $b=-1.729$, that forces an asymptote at $t=1$ (vertical black dashed line). The very good agreement between the fitting curve and the numerical results for the 11 squares chain seems to validate our assumption of an asymptote at $t=1$. 
The reason for an asymptote for this particular $t$ value is related to the energy dispersion of the TB model, where $t=1$ is the transition point where a gap opens between a lower flat band and an upper itinerant band.

%%%%%%%%%%%%%%%%%%%%%%%%%%%%%%%%%%%%%%%%%%%%%%%%%%%%%%%%%%%%%%%%%%%%%%%%%%%%%%%%%%% FIGURE %%%%%%%%%%%%%%%%%%%%%%%%%%%%%%%%%%%%%%%%%%%%%%%%%%%%%%%%%%%%%%%%%%%%%%%%%%%%%%%%%%%%%%%%%%%%%%%%%%%%%
\begin{figure}[h]
\begin{center}
\includegraphics[width=0.48 \textwidth ,height=0.25 \textheight]{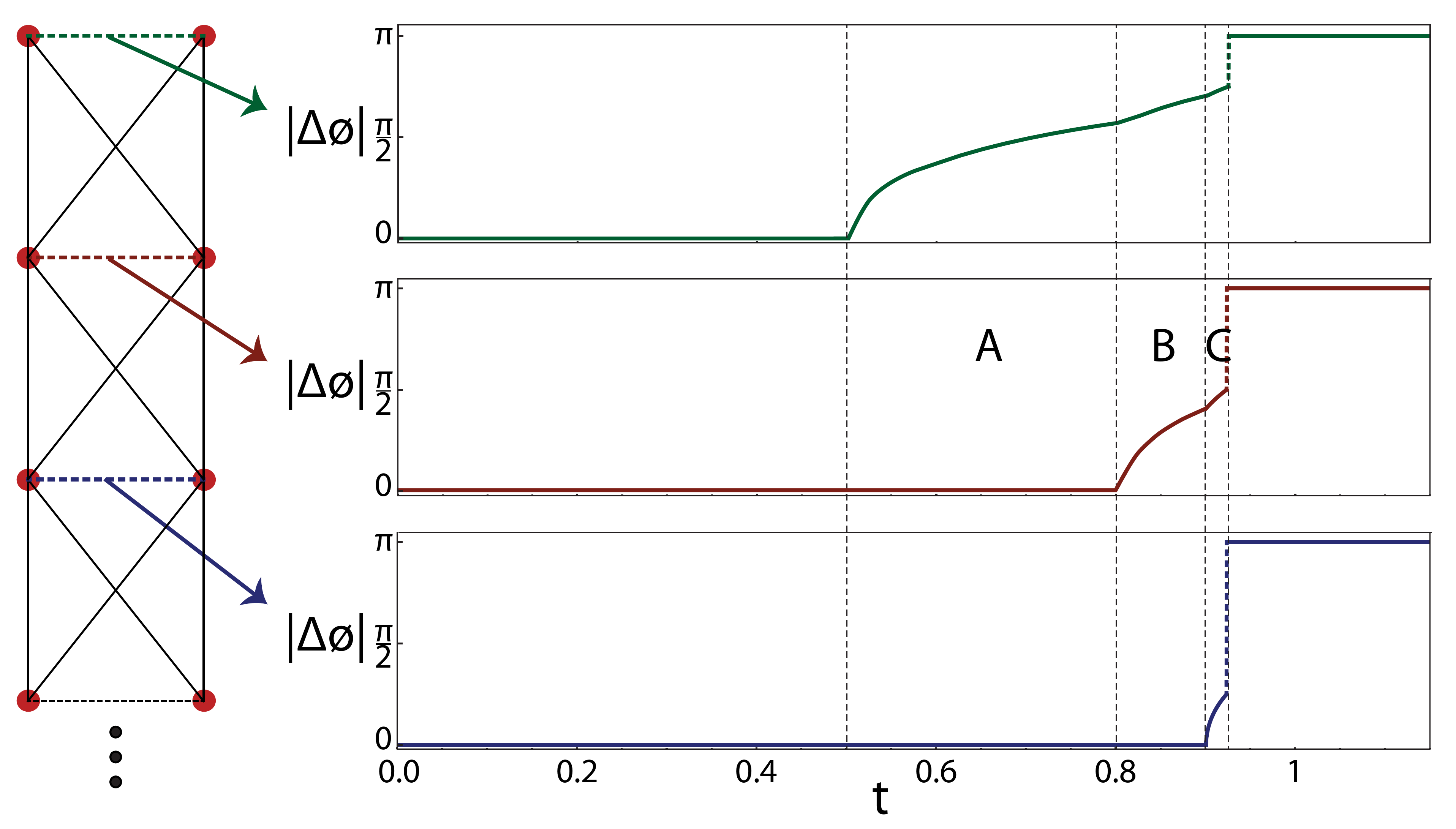}
\end{center}
\caption{Absolute value of the superconducting phase difference across the intra-rung junctions indicated by the arrows as a function of $t$. Only half of the six squares chain considered here is depicted, since the behavior of the other half is given by a reflection on the center of the chain. Regions A, B and C, delimited by the vertical dashed lines, are the same as in Fig.~\ref{fig:N_t_phase_diagram}. At the end of region C, all intra-rung phase differences jump discontinuously to $\pi$.}
\label{fig:phases_plot}
\end{figure}
Even though the range covered by the edge currents changes discontinuously, at the transitions between regions in Fig.~\ref{fig:N_t_phase_diagram}, the current flowing in each junction [directly related to the superconducting phase difference across that junction, see (\ref{eq:current_inter}) and (\ref{eq:current_intra})], varies continuously, except at the $t$ value above which edge currents disappear. 
This is shown for a chain with six squares in Fig.~\ref{fig:phases_plot}. 
When the phase difference across one of the rungs is neither 0 nor $\pi$, both the rung and the four junctions connected to it from below, in Fig.~\ref{fig:phases_plot}, carry currents, that is, the phase difference across these junctions, in absolute value, show the same behavior as the respective rung. 
As one goes from region A to region B, the second outermost rung develops a finite phase difference, meaning that the edge currents, which were already present at the outermost square, penetrate further into second outermost square.
Then, when one goes from region B to region C, the third outermost rung also develops a finite phase difference, and the edge currents penetrate into the respective square.
At the right end of region C, the phase differences of the rungs jump to $\pi$, \textit{i.e.}, the edge currents disappear from the chain, in agreement with Fig.~\ref{fig:N_t_phase_diagram}.
The chiral nature of the edge currents implies that the JJ chain state remains the same under an overall inversion of the currents at one or both of the edges. This amounts to a four-fold degeneracy for each JJ chain state with edge currents. 
One needs only to take the absolute values of the phase differences in the rungs, as in Fig.~\ref{fig:phases_plot}, to infer the presence and extent of the edge currents.

Our main case of study hereafter will be an open ladder with $N=7$ unit cells or, more exactly, rungs with $\pi$-junctions (6 squares). Figs.~\ref{fig:bandscurrentsplot}(b-f) show the decomposition of $\vert \psi \rangle$ in terms of the amplitudes of the $\alpha_{k_\pm}=\vert \alpha_{k_\pm} \vert e^{i\theta_{\alpha_{k_\pm}}}$ coefficients for five values of $t$ ($t=0.45,0.55,0.85,0.92,1.2$), along with the respective current configuration. %For each value of $t$ the energy dispersion of both bands in the continuum limit is represented and, superimposed on them, the amplitude $\vert\alpha_{k_\pm}\vert$ of the coefficients is indicated at the allowed $k$-states ($k=\frac{n\pi}{8}$ with $n=1,2,\dots,7$) according to the following notation: open orange circles indicate zero amplitude for the corresponding coefficients, meaning that they do not participate in the composition of $\vert \psi \rangle$ in (\ref{eq:psi_tb}), and solid black circles indicate occupied $k$-states with the relative value of the corresponding amplitude being given by its size (a bigger solid black circle has a higher amplitude of its coefficient and so the more weight it carries in the composition of $\vert \psi \rangle$). When a solid black circle is so small that it becomes imperceptible, which is the case, for instance, of $\vert\alpha_{k_+}\vert$ for $k=\frac{3\pi}{8},\frac{5\pi}{8},\frac{7\pi}{8}$ in curve $E_+$ of Fig.~\ref{fig:bandscurrentsplot}(e), it means that the amplitude of the coefficients is very small, almost negligible, but \textit{not zero}, in which case it would be an open orange circle. 
We see that, with the exception of the case of Fig.~\ref{fig:bandscurrentsplot}(f) where $\vert \psi \rangle$ is exclusively decomposed into $k$-states of the flat band, only $k$-states with odd $n$ have finite $\vert\alpha_{k_\pm}\vert$ [see Figs.~\ref{fig:bandscurrentsplot}(b-e)]. 
This can be explained if one recalls that odd $n$ harmonics are symmetric with respect to the center of the chain, whereas even $n$ harmonics are anti-symmetric. 
Since the numerically obtained minimum energy solutions for $\vert \psi \rangle$ in the chains are also symmetric with respect to their center, it follows that the decomposition of $\vert \psi \rangle$ will be such that all $\vert\alpha_{k_\pm}\vert$ in $k$-states with even $n$ are zero or, conversely, only $\vert\alpha_{k_\pm}\vert$ in $k$-states with odd $n$ can be finite. 

%%%%%%%%%%%%%%%%%%%%%%%%%%%%%%%%%%%%%%%%%%%%%%%%%%%%%%%%%%%%%%%%%%%%%%%%%%%%%%%%%%% FIGURE %%%%%%%%%%%%%%%%%%%%%%%%%%%%%%%%%%%%%%%%%%%%%%%%%%%%%%%%%%%%%%%%%%%%%%%%%%%%%%%%%%%%%%%%%%%%%%%%%%%%%
\begin{figure*}[t]
\begin{center}
\includegraphics[width=1 \textwidth]{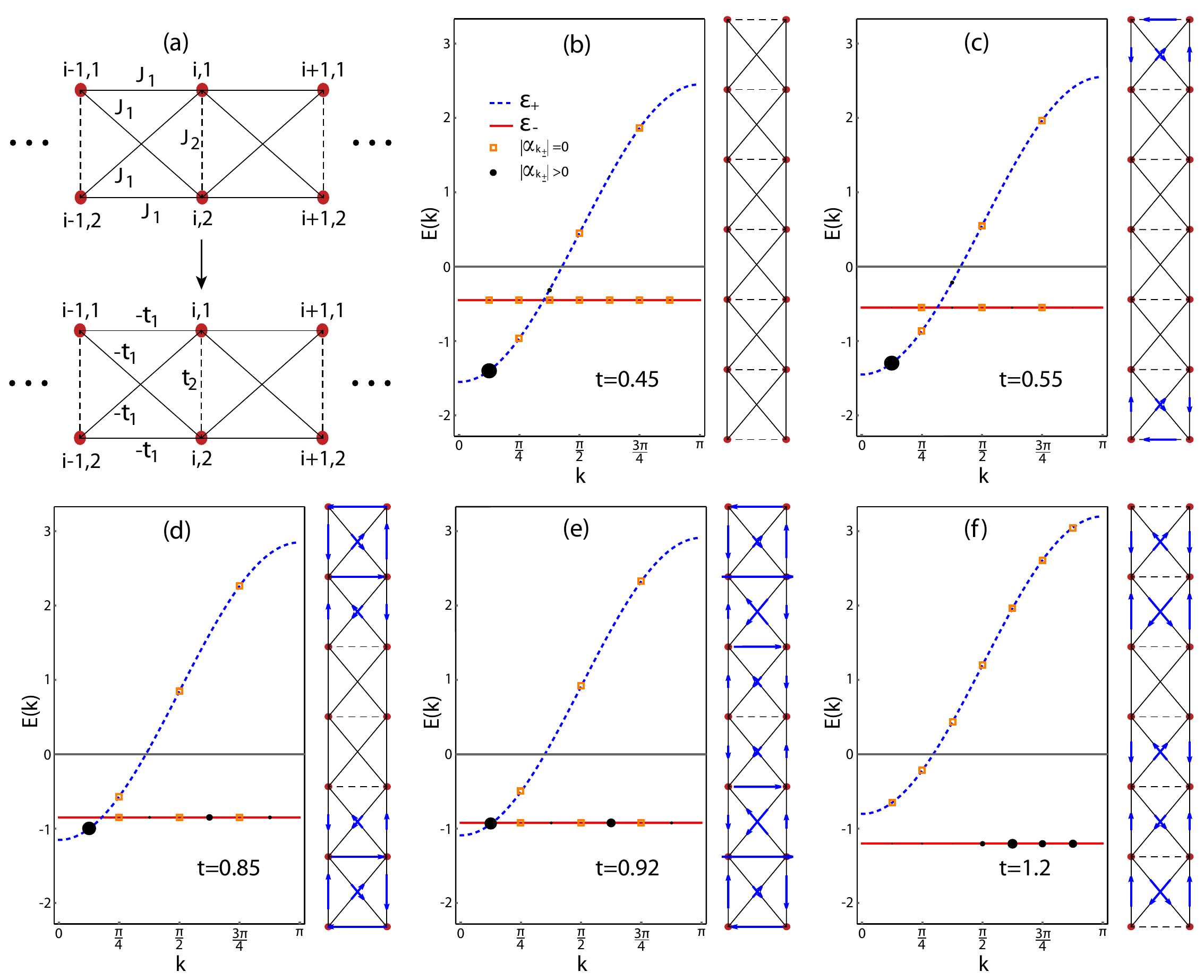}
\end{center}
\caption{(a) Correspondence between a two-leg ladder of JJs and a tight-binding system. 
%A factor of 1/2, present in (\ref{eq:hamilt_tb}), was omitted here in the substitutions $J_1 \to -t_1/2$ and $J_2 \to t_2/2$.
(b)-(f) Energy dispersion of the tight-binding bands, one itinerant (dashed blue curve $\varepsilon_+$) and one flat (solid red line $\varepsilon_-$) for five increasing values of $t$. 
Superimposed on the bands, the amplitude $\vert\alpha_{k_\pm}\vert$ of the coefficients is indicated at the allowed $k$-states ($k=\frac{n\pi}{8}$ with $n=1,2,\dots,7$) according to the following notation: open orange squares indicate zero amplitude for the coefficients, meaning that they do not participate in the composition of $\vert \psi \rangle$ in (\ref{eq:psi_tb}), and solid black circles indicate occupied $k$-states with the relative value of their amplitude being given by its size. When a solid black circle is so small that it becomes imperceptible, which is the case, for instance, of $\vert\alpha_{k_+}\vert$ for $k=\frac{3\pi}{8},\frac{5\pi}{8},\frac{7\pi}{8}$ in curve $\varepsilon_+$ of Fig.~\ref{fig:bandscurrentsplot}(e), it means that the amplitude of the coefficients is very small, almost negligible, but \textit{not zero}. The respective currents configuration is shown at the right of each plot.}
\label{fig:bandscurrentsplot}
\end{figure*}
For $t=0.45$ in Fig.~\ref{fig:bandscurrentsplot}(b), the $k$-states of the flat band $\varepsilon_-$ (solid red line) do not participate in $\vert \psi \rangle$, the superconducting phases  $\phi_{i,l}$ are the same for every site and the system exhibits no currents (one should remember that $t$ controls the relative strengths of $t_1$ and $t_2$, and a small $t_2$ does not make the system sufficiently frustrated so as to induce a chiral solution for the phases configuration that yields currents, as will be the case for higher values of $t$).

Vertical and diagonal inter-rung JJs, in the chain representation of Figs.~\ref{fig:bandscurrentsplot}(b)-(f), favor a zero phase difference across the junctions in the energy minimization and, conversely, horizontal JJs favor a $\pi$ phase difference [notice the opposite signs in the $t_1$ and $t_2$ terms in (\ref{eq:hamilt_tb})]. 
With increasing $t$, the effect of frustration eventually becomes strong enough so as to impose a compromise between these opposite tendencies and originate chiral solutions where some phase differences depart from 0 or $\pi$. 
Also, since the sites of the edge rungs of the ladder, relative to the bulk rungs, have less vertical and diagonal junctions "forcing" a zero phase difference, it is expected that chiral solutions should start to manifest at the edges first in the form of edge currents, which subsequently propagate to the bulk of the chain with increasing $t$ until complete coverage is reached, as can be seen in the evolution of the currents configuration in Figs.~\ref{fig:bandscurrentsplot}(c-e). But let us analyze each case separately for clarity.

Edge currents start to appear for $t>0.5$ ($t_2>t_1$). A little above this critical point, in Fig.~\ref{fig:bandscurrentsplot}(c), we can see the emergence of chiral currents at the edge squares, which coincide with the appearance of a small contribution in $k$-states with odd $n$ of the flat band (solid red line $\varepsilon_-$) that grows with $t$. 
For $t=0.55$ in Fig.~\ref{fig:bandscurrentsplot}(c) the chain is in region A of Fig.~\ref{fig:N_t_phase_diagram}. In this system edge currents are only present when $\vert\psi\rangle$ is composed by mixed contributions from $k$-states in both bands.

With increasing $t$, the magnitude of the inner currents of the outermost squares continuously increases, which in turn relax more and more the constraint they put on a zero phase difference between the sites of the second outermost rung. 
On the other hand, since the strength of $t_2$, relative to $t_1$, is increasing, a non-zero phase difference between the sites of this rung becomes more energetically favorable. 
From the combination of these two factors, above some $t$ value (more precisely above $t=0.809$), a compromise is reached and the chiral solution propagates into the second outermost squares as well, translating into an extension of the edge currents. 
This can be seen in Fig.~\ref{fig:bandscurrentsplot}(d), where the edge currents flow now within the two outermost squares of each end of the chain (the system is inside region B of the phase diagram of Fig.~\ref{fig:N_t_phase_diagram}). 
The $k$-states with odd $n$ in the flat band gain more weight in the decomposition of $\vert \psi \rangle$, following the tendency started in Fig.\ref{fig:bandscurrentsplot}(c).

The process is now repeated: the currents in the second outermost squares increase continuously with $t$ and, since this also means that $t_2$ is getting higher, a smaller increment in $t$ is required in order to make the chiral solution propagate into the third outermost squares, so now, for $t=0.92$ in Fig.\ref{fig:bandscurrentsplot}(e) (inside region C of the red dashed curve in Fig.~\ref{fig:N_t_phase_diagram}), the edge currents cover the whole ladder, starting at both ends and penetrating into the center.  The increments in $t$ required to propagate the edge currents to adjacent squares get smaller as $t$ is increased, which is why the width of the plateaus in the phase diagram of Fig.~\ref{fig:N_t_phase_diagram} decreases as t is increased.

Note once more that while the range of the edge currents changes discontinuously as they propagate to more and more squares, the values of the currents in each junction evolve continuously with $t$ (see Fig.~\ref{fig:phases_plot}). 
In this $t$ evolution, briefly illustrated by the sequence in Figs.~\ref{fig:bandscurrentsplot}(c-e), $k$-states with odd $n$ in the flat band $\varepsilon_-$ (solid red line) become increasingly dominant, in relation to the itinerant band $\varepsilon_+$ (dashed blue curve), in the decomposition of $\vert\psi\rangle$. This behavior is expected because, as $t$ increases, the energy of the flat band lowers and, simultaneously, the energy of the itinerant band grows, so $\vert\psi\rangle$ will transfer the weight in $\vert\alpha_{k_+}\vert$ to $\vert\alpha_{k_-}\vert$ as much as possible in order to minimize its energy. 

%%%%%%%%%%%%%%%%%%%%%%%%%%%%%%%%%%%%%%%%%%%%%%%%%%%%%%%%%%%%%%%%%%%%%%%%%%%%%%%%%%% FIGURE %%%%%%%%%%%%%%%%%%%%%%%%%%%%%%%%%%%%%%%%%%%%%%%%%%%%%%%%%%%%%%%%%%%%%%%%%%%%%%%%%%%%%%%%%%%%%%%%%%%%%
\begin{figure}[h]
\begin{center}
\includegraphics[width=0.48 \textwidth ,height=0.275 \textheight]{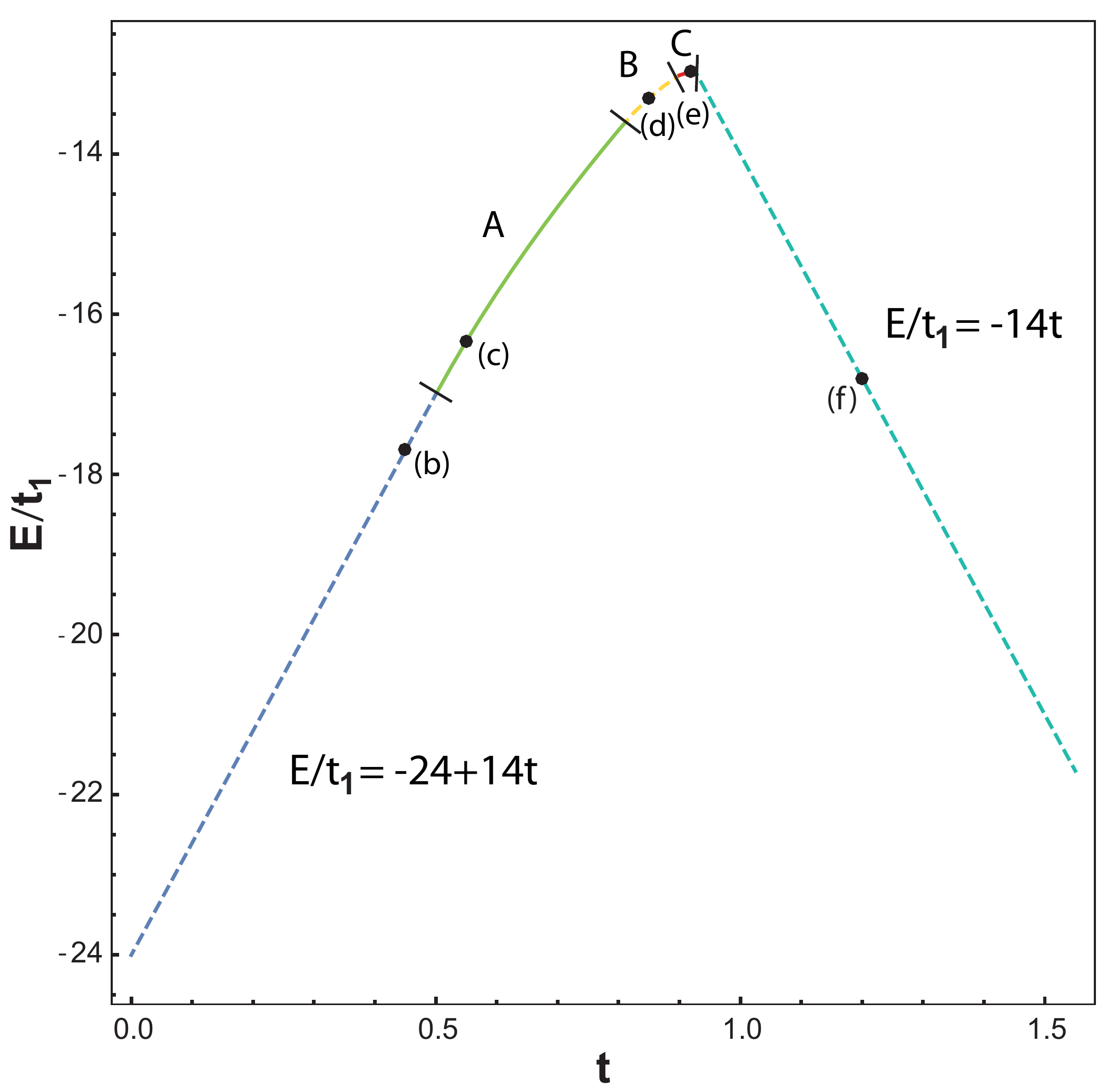}
\end{center}
\caption{Energy, normalized by the hopping constant $t_1$, as a function of $t=\frac{t_2}{2t_1}$ for the chain with 6 squares, depicted at the top of Fig.~\ref{fig:N_t_phase_diagram}. Transitions between different regions are marked by a change in color and curve style, which alternates between dashed and solid. Regions A, B, and C are the same as those of the red dashed curve in Fig.~\ref{fig:N_t_phase_diagram}. Points labeled (b) through (f) give the energy of the corresponding cases in Fig.~\ref{fig:bandscurrentsplot}. The explicit expressions for the energy in the two linear regions is shown.}
\label{fig:energy}
\end{figure}
Above the upper limit for $t$ where the maximum range for these chiral edge currents is reached, as in Fig.~\ref{fig:bandscurrentsplot}(f), where $t=1.2$ (above the sudden drop in the red dashed curve in Fig.~\ref{fig:N_t_phase_diagram}), the superconducting state no longer supports chiral edge currents, whose existence always involves at least one current flowing in a rung, which is not the case now [there are no horizontal currents in Fig.~\ref{fig:bandscurrentsplot}(f)]. 

The persistence of zigzag current loops within the squares has a distinct origin from that of the edge currents in Figs.~\ref{fig:bandscurrentsplot}(b-e). 
Indeed, one sees in Fig.~\ref{fig:bandscurrentsplot}(f) that $\vert\psi\rangle$ is decomposed in a combination of all degenerate and localized $k$-states of the flat band $\varepsilon_-$ with different relative amplitudes, which turn out to be arbitrary (within the limits set by the norm of the JJ array state, $\vert \psi \vert=\sqrt{2N}$), and no $k$-states of the itinerant band $\varepsilon_+$ participate in $\vert\psi\rangle$. 
Additionally, one sees that this is the only case were coefficients from $k$-states with even $n$ also appear in the decomposition of $\vert\psi\rangle$, which is only possible because they correspond to localized states that are effectively decoupled from each other, so that the global symmetry of the chain becomes a set of 6 local symmetry conditions (one for each square). 
The localized states correspond to ``dimerized'' rungs, independent of each other and with fixed intra-rung phase differences $\Delta\phi=\pi$ [horizontal phase differences in Fig.~\ref{fig:phases_plot} and Fig.~\ref{fig:bandscurrentsplot}(f) are all $\pi$ for $t=1.2$]. 
But, since these localized ``dimers'' are independent, vertical phase differences between adjacent rungs can take arbitrary values, and these arbitrary vertical phase differences may or may not produce inner currents in the squares, as seen in Fig.\ref{fig:bandscurrentsplot}(f). 
The energy contribution of the four inner currents in a square cancel out by pairs [two currents flowing in opposite directions in a square of Fig.~\ref{fig:bandscurrentsplot}(f) have a $\pi$ phase difference which cancel the cosines in $H_1$ in (\ref{eq:hamiltonian_JJ1})], so they have no effect on the total energy of $\vert\psi\rangle$.

%%%%%%%%%%%%%%%%%%%%%%%%%%%%%%%%%%%%%%%%%%%%%%%%%%%%%%%%%%%%%%%%%%%%%%%%%%%%%%%%%%% FIGURE %%%%%%%%%%%%%%%%%%%%%%%%%%%%%%%%%%%%%%%%%%%%%%%%%%%%%%%%%%%%%%%%%%%%%%%%%%%%%%%%%%%%%%%%%%%%%%%%%%%%%
\begin{figure*}[t]
\begin{center}
\includegraphics[width=1 \textwidth]{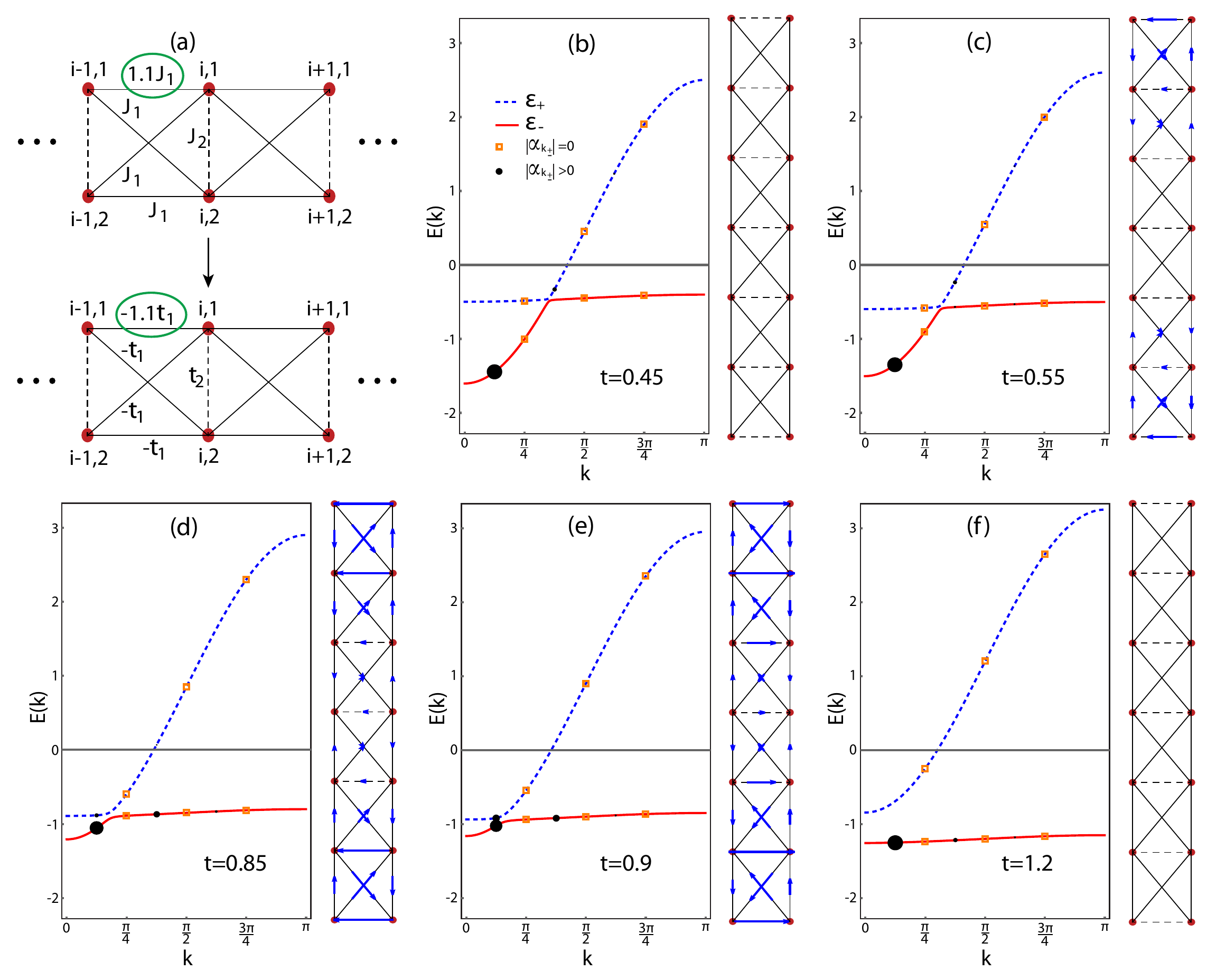}
\end{center}
\caption{Same as in Fig.~\ref{fig:bandscurrentsplot}. The modification in the value of the top $t_1$ is highlighted in (a). Because of this modification, both bands in (b)-(f) are now itinerant, but the red solid band $\varepsilon_-$ becomes quasi-flat for $t>1$, as in (f). Bands $\varepsilon_-$ and $\varepsilon_+$ do not intersect [even in (b) and (c)].}
\label{fig:bandscurrentsplotiten}
\end{figure*}
Above $t=1$, a gap opens between $\varepsilon_+$ and $\varepsilon_-$. When the bands separate, our numerical results show that edge currents are not present, regardless of chain size. 
This is the reason why in the $N_{ec}$ vs. $t$ phase diagram of Fig.~\ref{fig:N_t_phase_diagram} there is an asymptote (vertical black dashed line) at $t=1$ limiting the fitting curve for an infinite-sized chain (green dashed curve). 
The fact that, for chains with 6 and 11 squares, states with edge currents are destroyed below $t=1$ is a consequence of their finite size. 
The longer chain, the closer the upper $t$ limit for the presence of edge currents gets to $t=1$.   

The evolution of the 6 squares chain energy with $t$ is shown is Fig.~\ref{fig:energy}. 
Linear behavior is observed if edge currents are absent. 
For $t<0.5$ (dark blue dashed curve), there are no finite superconducting phase differences ($\Delta\phi=0$) since every site has the same phase. So $t_1$ hoppings lower the energy [$-t_1\cos(\Delta\phi)$] while $t_2$ hoppings increase it [$t_2\cos(\Delta\phi)$], and the energy is given by $E=-24t_1+7t_2$, with 24 being the total number of $t_1$ hoppings in the 6 squares (four per square) of the chain and 7 is the number of $t_2$ hoppings (one per rung).

For $t\gtrsim 0.97$ (light blue dashed curve to the right of region C in Fig.~\ref{fig:energy}), the superconducting phase difference between the two sites of each rung is $\pi$ [the energy is lowered by the sum of all seven individual contributions $t_2\cos(\pi)$ of the $t_2$ hoppings] and, as we explained before, superconducting phase differences between the pairs of sites in adjacent rungs are arbitrary, but the energy contributions of the four inter-rung $t_1$ hoppings cancel in pairs [case of Fig.~\ref{fig:bandscurrentsplot}(f)], so the total energy is just $E=-7t_2$. 

As we go from region A to region B and then to region C with increasing $t$ in Fig.~\ref{fig:energy}, where superconducting phase differences in junctions that carry currents are neither 0 nor $\pi$, and since the energy contributions of the junctions do not cancel each other (as in the region given by the light blue dashed curve for $t>1$, for the reasons stated above), we see that the energy profile deviates more and more from a linear behavior due to the edge currents that penetrate further and further into the ladder.

\subsubsection{Ladder with a quasi-flat band}

In this subsection we will briefly analyze the consequences produced by a small change in the value of one of the $t_1$ hoppings of the squares, as illustrated in Fig.~\ref{fig:bandscurrentsplotiten}(a). This modification is made with the objective of explicitly broadening the flat band $\varepsilon_-$ of Fig.~\ref{fig:bandscurrentsplot}, turning it into another itinerant band (or quasi-flat band for high $t$ values). 
Note that with this modification bands $\varepsilon_-$ and $\varepsilon_+$ do no intersect for any $t$ value, as seen in Figs.~\ref{fig:bandscurrentsplotiten}(b)-(f).

In this case $\vert \psi \rangle$ is always decomposed in mixed contributions from the $k$-states with odd $n$ of both bands, even though these contributions are small or almost zero for $\varepsilon_+$ [see blue dashed curves in Figs.~\ref{fig:bandscurrentsplotiten}(b)-(f)], regardless of the $t$ value. 
For $t\lesssim 0.5$, as in Fig.~\ref{fig:bandscurrentsplotiten}(b), all sites have again the same superconducting phase and there are no edge currents. When $t$ is above this critical point ($t\gtrsim 0.5$) edge currents appear and propagate faster than in the system of Fig.~\ref{fig:bandscurrentsplot}. 
For $t=0.55$ in Fig.~\ref{fig:bandscurrentsplotiten}(c), edge currents already occupy the two outermost squares at each end, as opposed to occupying only the outermost one as in Fig.\ref{fig:bandscurrentsplot}(c).
 When $t=0.85$, in Fig.~\ref{fig:bandscurrentsplotiten}(d), the edge currents have already propagated to the whole chain. The same behavior is seen in Fig.~\ref{fig:bandscurrentsplotiten}(e), but with higher currents in the middle squares. 
%Notice also that in Figs.~\ref{fig:bandscurrentsplotiten}(c)-(e) the four inner currents in the squares with edge currents do not have the same absolute value, since the modification of the value of one $t_1$ hopping [see Fig.~\ref{fig:bandscurrentsplotiten}(a)] disrupts the balance between the four $t_1$ hoppings.

Once more, as $t$ is increased and a gap opens between the bands of the TB model the chain no longer supports edge currents. This is observed in Fig.~\ref{fig:bandscurrentsplotiten}(f), where $t=1.2$. There is, however, a noticeable difference between the current configurations of Fig.~\ref{fig:bandscurrentsplot}(f) and Fig.~\ref{fig:bandscurrentsplotiten}(f): the arbitrary currents flowing in the four inner $t_1$ hoppings of the squares seen in the former are absent in the latter. In Fig.~\ref{fig:bandscurrentsplot}(f), arbitrary inner currents appeared because $\vert \psi\rangle$ was decomposed only in localized states of the flat band, whereas in Fig.~\ref{fig:bandscurrentsplotiten}(f), $\vert\psi\rangle$ is almost only written as a linear combination of $k$-states with odd $n$ of the solid red band $\varepsilon_-$, which is quasi-flat. Therefore, there are no localized states, so arbitrary currents inside the squares are absent (even if $\vert \psi \rangle$ was exclusively decomposed in the quasi-flat band), in the chain of Fig.~\ref{fig:bandscurrentsplotiten}(a).

The analysis of the system of Fig.~\ref{fig:bandscurrentsplotiten}, when contrasted with that of the system of Fig.~\ref{fig:bandscurrentsplot}, allows one to draw two important conclusions: the absence of edge currents in these types of chains does not imply that $\vert\psi\rangle$ is decomposed only in $k$-states of one of the bands, as is the case in Figs.\ref{fig:bandscurrentsplot}(b) and (f), since there are also no edge currents in Figs.\ref{fig:bandscurrentsplotiten}(b) and (f), and $\vert\psi\rangle$, in both these cases, is decomposed in mixed contributions of $k$-states with odd $n$ in both bands (even though they are small in the blue dashed band $\varepsilon_+$). This implication seems only valid in the case where a flat band is present. 

The other conclusion is that there is no qualitative difference in the decomposition of $\vert \psi \rangle$ between Figs.~\ref{fig:bandscurrentsplotiten}(e) and (f), and yet the edge currents disappear from one to the other, which means that the appearance of edge currents are not to be explained by qualitative changes in the decomposition of  $\vert \psi \rangle$. 
What changes from Fig.~\ref{fig:bandscurrentsplotiten}(e) to Fig.~\ref{fig:bandscurrentsplotiten}(f) is that a gap opens between the bands of the TB system. The corresponding disappearance of edge currents, also observed in Figs.~\ref{fig:bandscurrentsplot}(e) and (f), seems to indicate that it is related to the opening of the gap and may follow from some topological argument.

%%%%%%%%%%%%%%%%%%%%%%%%%%%%%%%%%%%%%%%%%%%%%%%%%%%%%%%%%%%%%%%%%%%%%%%%%%%%%%%%%%% SECTION %%%%%%%%%%%%%%%%%%%%%%%%%%%%%%%%%%%%%%%%%%%%%%%%%%%%%%%%%%%%%%%%%%%%%%%%%%%%%%%%%%%%%%%%%%%%%%%%%%%%
\section{Conclusions}
\label{sec:conclusions}

In this paper, we considered a frustrated JJ ladder with diagonal couplings and open boundary conditions. We found they exhibit new phenomena, in particular the emergence of what we designated as chiral edge currents. %with different penetration depths when some superconducting phase differences between adjacent sites at the edges are neither 0 nor integer multiples of $\pi$, 
Upon deriving the correspondence between the JJ chain and an analogous TB model, we showed how one can write the state $\vert\psi\rangle$ of the JJ chain in terms of the allowed $k$-states of the TB bands [see (\ref{eq:psi_tb})]. 
The energy of the chain can then be calculated as the expectation value of the TB Hamiltonian [see (\ref{eq:energy_tb})].

The appearance and propagation of edge currents from the outermost squares to the bulk of the ladder as a function of $t$
%, the relative strength of the hopping constants in the TB model, which corresponds to the relative strength of the Josephson couplings in the JJ chain, 
is presented in the phase diagram of Fig.~\ref{fig:N_t_phase_diagram}. 
With increasing $t$, the edge currents disappear only after they reach complete coverage of the whole ladder (except for the central square in chains with odd number of squares, for symmetry reasons). 
Since in ladders with larger sizes the edge currents take longer to reach complete coverage, the upper limit of $t$ for the existence of edge currents changes with ladder size. 
In the limit where the ladder length goes to infinity, this upper limiting $t$ is shown, from a fitting curve that agrees very well with our numerical results for a finite-size ladder with 11 squares, to tend asymptotically to $t=1$. 
This point is special in the sense that it corresponds, in the TB model, to the point where a gap opens between the TB bands, which seems to suggest that the appearance/disappearance of edge currents in this and similar frustrated Q1D arrays may in the future  be predicted as a consequence of some topological argument.

The decomposition of the JJ ladder state $\vert\psi\rangle$ in the $k$-states of the TB bands was carefully analyzed for different $t$ values, corresponding to different currents configurations, in the case of the ladder with 6 squares in Fig.~\ref{fig:bandscurrentsplot}. 
With increasing $t$, $\vert\psi\rangle$ progressively goes from a linear combination of only $k$-states in the itinerant band to a linear combination of only $k$-states in the flat band, when a gap opens between the bands. 
In between these two limits, for the $t$ values where states with edge currents exist, $\vert\psi\rangle$ has mixed contributions from both bands. 
For $t>1$, $\vert\psi\rangle$ is a superposition of localized states of the flat band, which reflect the existence of independent ``dimerized'' rungs with a fixed $\pi$ superconducting phase difference. 
Their independence translates into arbitrary phase differences between adjacent rungs, which in turn lead to the appearance of inner currents in the squares of the ladder [see Fig.~\ref{fig:bandscurrentsplot}(f)], although of a different nature from the chiral edge currents.

Finally, we studied the same 6 squares ladder as before, but modified slightly the strength of one of the hoppings in the unit cell [see Fig.~\ref{fig:bandscurrentsplotiten}(a)], therefore turning the flat band into another itinerant (quasi-flat for high $t$ values) band. 
Two main changes in the behavior of the system were observed: for all $t$ values, $\vert\psi\rangle$ is a linear combination of $k$-states of both bands; additionally, for $t>1$, $\vert\psi\rangle$ is no longer a linear combination of localized states, so the inner currents present in Fig.~\ref{fig:bandscurrentsplot}(f) are absent in Fig.~\ref{fig:bandscurrentsplotiten}(f).

The existence of the edge current states addressed in this paper poses an interesting question, whose answer is left open for future work: motivated by the bulk-edge correspondence stated in the introduction, are these edge current states a consequence of some topological non-trivial invariant of the bulk of the corresponding TB system, in the $t$ range predicted by our numerical results? Note that the respective TB system is topologically trivial with respect to the winding number, but the JJ array states in the classical limit map into a subspace of states of the TB Hamiltonian, and this restriction could generate some non-obvious topological invariant. Or should this topological invariant, if it can indeed be defined, as our analysis seems to indicate, be searched elsewhere, perhaps not directly associated to the  JJ $\leftrightarrow$ TB correspondence?

Recently, a different kind of chiral currents was found in optical lattices of cold-atoms with synthetic gauge fields \cite{Dhar2013,Celi2014,Atala2014,Greschner2015}. 
The behavior of these chiral currents when varying the ratio of the hopping constants of the optical ladders resembles that of type-II superconductors, given that it is also observed a transition between a ``Meissner'' phase, where bosonic currents circulate around the ladder to counteract the effect of the magnetic flux, and an ``Abrikosov'' vortex phase, where a vortex structure of circulating bulk currents spreads along the ladder.
Even though the chiral edge currents characterized in this paper have a different origin than those observed in optical lattices thus far, their observation in these kinds of lattices, if built with a frustrated geometry similar to the one studied here, should not be ruled out.

%%%%%%%%%%%%%%%%%%%%%%%%%%%%%%%%%%%%%%%%%%%%%%%%%%%%%%%%%%%%%%%%%%%%%%%%%%%%%%%%%%% FIGURE %%%%%%%%%%%%%%%%%%%%%%%%%%%%%%%%%%%%%%%%%%%%%%%%%%%%%%%%%%%%%%%%%%%%%%%%%%%%%%%%%%%%%%%%%%%%%%%%%%%%%
\begin{figure}[h]
\begin{center}
\includegraphics[width=0.35 \textwidth ,height=0.17 \textheight]{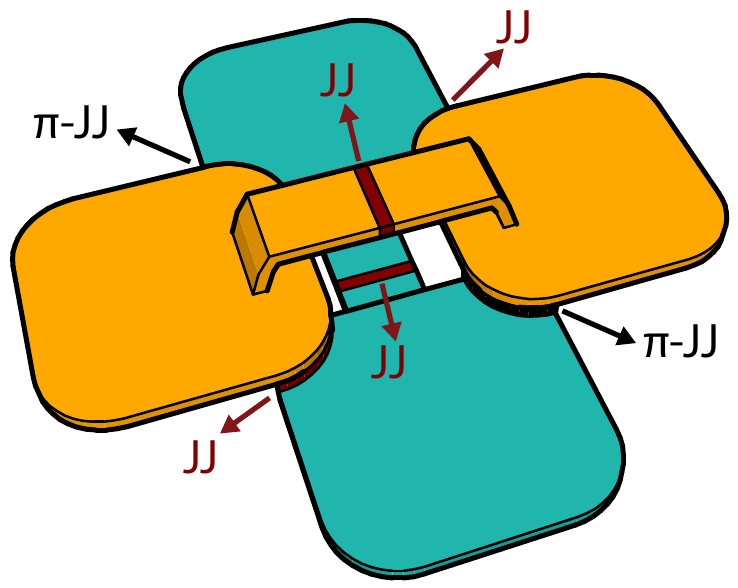}
\end{center}
\caption{Scheme of a plaquette with our proposed frustrated geometry. 
Red and black layers signal, respectively, typical junctions and $\pi$-junctions between superconductors, as indicated by the arrows. 
The upper (yellow) and lower (blue) crossed junctions are depicted as ``bridges'' for visual clarity, but can be planar.
A JJ ladder is constructed by longitudinal repetitions of this plaquette.}
\label{fig:plaquette}
\end{figure}
Several experimental groups have shown to be able to manufacture and manipulate JJ arrays with different geometries \cite{Andersson2003,Takahide2006,Pop2008,Gladchenko2009,Pop2010,Bell2014,Vogt2015}. The particular geometry considered here, because of its intrinsic frustration, imposes that the rungs should consist of sign-reversed two-band superconductors or, more feasibly, interconnected $\pi$-junctions \cite{Bauer2004,Kirtley2005,Hilgenkamp2008,Gurlich2010}. 
A scheme of a plaquette, the fundamental building block of the JJ ladders, is depicted in Fig.~\ref{fig:plaquette}.
This scheme follows closely the one experimentally realized by Binder \textit{et al.} \cite{Binder2000} [see Fig.~1(b) of this reference], apart from the replacement of two of the junctions with $\pi$-junctions and the introduction of the two crossed JJs shown in Fig.~\ref{fig:plaquette}.
The four red junctions in Fig.~\ref{fig:plaquette} can represent, for instance, aluminum oxide insulating layers between niobium superconductors \cite{Binder2000} ($Nb$/$Al_2O_3$/$Nb$ type junctions).
It has been shown that the introduction of an additional ferromagnetic layer between the insulating layer and the superconductor can produce, above a critical thickness, a $\pi$-junction \cite{Weides2006,Sprungmann2009,Gingrich2016}.
The introduction of these ferromagnetic layers is one way of producing the two $\pi$-junctions represented by the black layers in Fig.~\ref{fig:plaquette}.
The thickness and temperature of both the insulating and ferromagnetic layers serves also as parameters that determine the strength of the Josephson couplings \cite{Kontos2002,Bannykh2009}.
Therefore, these parameters can be controlled and adjusted to yield the desired $J_2/J_1$ ratio.

Some preliminary results, concerning the extension of the present study to finite 2D JJ arrays, seem to show the same qualitative behavior as the one described in this paper for frustrated JJ ladders, that is, there is a $t$ value above which chiral edge currents appear and, with increasing $t$, these currents propagate from the surface to the center of the 2D JJ array.

%%%%%%%%%%%%%%%%%%%%%%%%%%%%%%%%%%%%%%%%%%%%%%%%%%%%%%%%%%%%%%%%%%%%%%%%%%%%%%%%%%% SECTION %%%%%%%%%%%%%%%%%%%%%%%%%%%%%%%%%%%%%%%%%%%%%%%%%%%%%%%%%%%%%%%%%%%%%%%%%%%%%%%%%%%%%%%%%%%%%%%%%%%%

\section*{Acknowledgments}\label{sec:acknowledments}

This work is funded by FEDER funds through the COMPETE 2020 Programme and National Funds throught FCT - Portuguese Foundation for Science and Technology under the project UID/CTM/50025/2013.
A.M.M. acknowledges the financial support from the FCT through the grant SFRH/PD/BD/108663/2015.
F.D.R.S acknowledges the financial support from the FCT through the grant BI-22/I3N-SET-2015.
R.G.D. thanks the support by the Beijing CSRC.

%%%%%%%%%%%%%%%%%%%%%%%%%%%%%%%%%%%%%%%%%%%%%%%%%%%%%%%%%%%%%%%%%%%%%%%%%%%%%%%%%%% SECTION %%%%%%%%%%%%%%%%%%%%%%%%%%%%%%%%%%%%%%%%%%%%%%%%%%%%%%%%%%%%%%%%%%%%%%%%%%%%%%%%%%%%%%%%%%%%%%%%%%%%

\bibliography{edgecurrents}

\end{document}